\documentclass{phep}       

\received{xx January 2025}
\published{xx March 2025}

\usepackage{amsmath}
\usepackage{amssymb}
\usepackage[hidelinks]{hyperref}

\def\be{\begin{equation}}
\def\ee{\end{equation}}
\def\bea{\begin{eqnarray}}
\def\eea{\end{eqnarray}}

\renewcommand{\thefootnote}{\fnsymbol{footnote}}

\begin{document}

\title{Probing the Electroweak Phase Transition in the Dark Sector}

\author{Maimoona Razzaq,\auno{1,2}\footnotemark\ Nico Benincasa, \auno{3} Luigi Delle Rose,\auno{1,2} and Luca Panizzi, \auno{1,2}}
\address{$^1$ Dipartimento di Fisica, Università della Calabria, Arcavacata di Rende, I-87036, Cosenza, Italy}
\address{$^2$ INFN, Gruppo Collegato di Cosenza, Arcavacata di Rende, I-87036, Cosenza, Italy}
\address{$^3$ School of Physics, University of Electronic Science and Technology of China, 611731 Chengdu, China}

\begin{abstract}
We study an extension of the Standard Model with a dark $SU(2)$ gauge group, where a dark scalar doublet interacts with the Standard Model Higgs through a portal coupling, inducing mixing after symmetry breaking. A custodial symmetry ensures the stability of the dark gauge bosons, making them viable dark matter candidates. Scanning the parameter space of the model under collider and astrophysical constraints, we find regions that yield the observed relic density and strong first-order phase transitions. The resulting gravitational-wave signals fall within the reach of upcoming space-based detectors.
\end{abstract}

\maketitle


\renewcommand{\thefootnote}{\fnsymbol{footnote}}
\footnotetext{Speaker and Corresponding author: maimoonarazzaq10@gmail.com}

\section{Introduction}
\label{intro}

A wealth of cosmological and astrophysical evidence points to the fact that the visible, baryonic component of matter represents only a small share of the total matter content of the Universe. 
The dominant fraction is composed of dark matter (DM), a non-baryonic form of matter that interacts mainly through gravity and remains unaffected by electromagnetic or strong interactions. 
Explaining the nature of this component has motivated numerous extensions of the Standard Model (SM), which introduce new stable particles weakly coupled to SM fields. 
Depending on the interaction strength and dynamics, these models can be tested through a combination of direct and indirect detection experiments, together with collider searches. 
In recent years, gravitational waves (GWs) have offered an additional avenue to probe such theories, particularly those involving extra scalar fields capable of generating first-order phase transitions (FOPTs) in the early Universe, thereby producing a stochastic GW background observable by future detectors, such as LISA~\cite{Caprini:2019egz}, DECIGO~\cite{Kawamura:2011zz}, BBO~\cite{Corbin:2005ny}, TianQin~\cite{TianQin:2015yph}, and Taiji~\cite{Hu:2017mde}.
In contrast, the electroweak phase transition predicted by the SM proceeds as a continuous crossover~\cite{Kajantie:1996mn}, suggesting that new fields or interactions beyond the SM are required to achieve a strong first-order transition capable of producing an observable GW signal.

Based on \cite{Benincasa:2024pfs,Benincasa:2025tdr}, this study investigates a minimal extension of the SM that naturally accommodates DM and predicts FOPTs capable of producing detectable gravitational waves.  To assess the viability of this scenario, we systematically explore its parameter space by applying a combination of theoretical, collider, electroweak precision, and astrophysical constraints.  We then trace the thermal evolution of the scalar potential to determine the conditions under which strong FOPTs occur and generate a stochastic GW background~\cite{Athron:2023xlk}, subsequently examining whether these signals could be observed by forthcoming space-based GW detectors.

\section{The Dark Matter Model}
\begin{figure}[htbp]
\centering
\includegraphics[width=.3\textwidth]{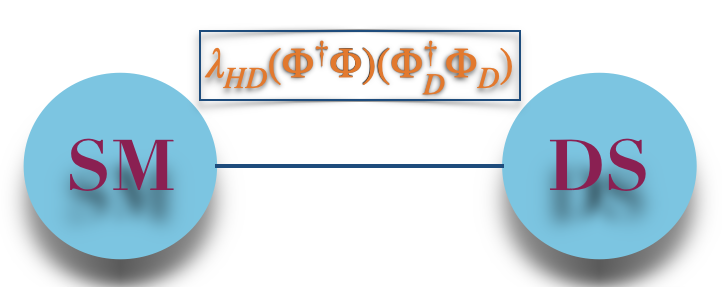}
\caption{\label{MS_DS}Interaction between SM and Dark Sector}
\end{figure}

This work explores a minimal extension of the SM first introduced in~\cite{Hambye:2008bq}, where an additional non-Abelian gauge group, $SU(2)_D$, is incorporated into a hidden (dark) sector. The motivation for such an extension arises from the ongoing search for viable DM candidates beyond the SM framework. Within this setup, the model naturally predicts stable dark vector bosons that serve as DM candidates. Their stability originates from a custodial symmetry emerging from the structure of the scalar sector. In what follows, we outline the essential theoretical ingredients of this model that are relevant for our analysis.\\
The Lagrangian describing the new sector can be written as:
\begin{align}
\mathcal{L} &= - \frac{1}{4} F_{D\mu\nu}^a F_D^{a\mu\nu} 
+ (D_\mu \Phi_D)^\dagger (D^\mu \Phi_D) 
+ \mu_D^2 (\Phi_D^\dagger \Phi_D)\notag \\[2mm]
&\quad - \lambda_D (\Phi_D^\dagger \Phi_D)^2 
- \lambda_{HD} (\Phi_D^\dagger \Phi_D)(\Phi^\dagger \Phi)\,,
\end{align}
where $F_D^{a\mu\nu}$ is the field-strength tensor associated with the dark gauge symmetry, and the covariant derivative takes the form:
\begin{equation}
D_\mu = \partial_\mu - i g_D T_D^a V_{D\mu}^a\,,
\end{equation}
with $T_D^a$ denoting the generators of $SU(2)_D$ in the fundamental representation. The term proportional to $\lambda_{HD}$ acts as a Higgs-portal interaction, coupling the SM Higgs doublet $\Phi$ to the dark scalar field $\Phi_D$ through a bilinear operator, as illustrated in Figure~\ref{MS_DS}.  

The Lagrangian is symmetric under an $SO(4)$ global symmetry, which is spontaneously broken to a custodial $SO(3)$ symmetry when the dark scalar acquires a vacuum expectation value (VEV) $v_D$. This custodial symmetry guarantees the stability of the dark vector bosons, preventing them from decaying into SM states and thus making them suitable DM candidates. Nevertheless, introducing an extended fermion sector may break this symmetry, lift the mass degeneracy among the dark gauge bosons at the loop level, and enable purely SM decay modes for one of them~\cite{Belyaev:2022zjx,Belyaev:2022shr}.

\section{Scanning the Parameter Space}

Although this model represents a minimal extension of the SM, it is subject to several complementary constraints. We first consider the theoretical requirements ensuring that the scalar potential is bounded from below and that perturbative unitarity is maintained. We then enforce cosmological and astrophysical limits associated with the DM candidate, as well as collider constraints restricting possible modifications of the scalar sector.

Numerically, we perform random scans over the four free parameters of the model: the masses of the dark gauge bosons and of the dark Higgs, $m_{V_D}$ and $m_{H_D}$, are varied logarithmically between 10~MeV and 100~TeV, $g_D$ logarithmically from $10^{-5}$ to $4\pi$, and $\cos\theta_S$, the mixing angle between the scalars induced by the portal coupling, linearly between 0 and 1. Additional focused scans are carried out to increase the density of viable points satisfying all constraints.

\begin{figure}[htbp]
\centering
\includegraphics[scale=0.57]
{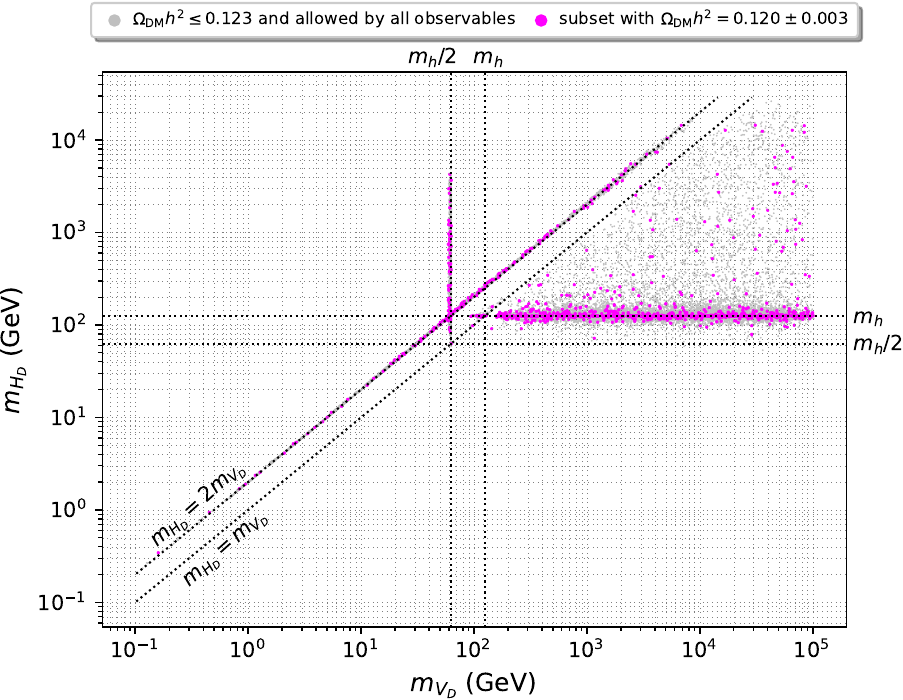}
\caption{Projections of the parameter space satisfying all constraints. Magenta points also reproduce the observed relic density within $3\sigma$, while grey points correspond to underabundant dark matter.}
\label{fig:summaryscatter}
\end{figure}

Figure~\ref{fig:summaryscatter} shows the allowed regions in the $\{m_{V_D}, m_{H_D}\}$ plane after applying the full set of constraints. Grey points correspond to parameter sets yielding an underabundant DM relic density, while magenta points reproduce the Planck 3$\sigma$ range, $\Omega_{\text{DM}}h^2 = 0.120 \pm 0.003$ \cite{Planck:2018vyg}. The region above the line $m_{H_D}=2m_{V_D}$ and for $m_{V_D} \lesssim m_h/2$ is excluded as it predicts a relic density exceeding the observed value, while the lower trapezoidal region is ruled out by direct detection experiments due to large DM--nucleon scattering cross-sections.

The distribution of viable points spans $m_h/2 \lesssim m_{H_D} \lesssim m_{V_D}$ and clusters around three characteristic regimes: $m_{H_D} \simeq m_h$, $m_{V_D} \simeq m_h/2$, and $m_{H_D} \simeq 2m_{V_D}$. The conditions $m_{V_D} \simeq m_h/2$ and $m_{H_D} \simeq 2m_{V_D}$ correspond to resonant regions where annihilation through an $s$-channel scalar is enhanced, effectively reducing the DM relic abundance even for small $g_D$. When the two scalars become nearly degenerate, $m_{H_D} \simeq m_h$, their mixing can increase, allowing both to decay into SM final states. This behaviour also emerges from collider constraints, particularly from \texttt{HiggsSignals}, which limit the scalar mixing to ensure that the SM-like Higgs couplings remain consistent with LHC observations.

\section{Key Parameters of the Phase Transition}

The properties of the GW signal are governed by a few macroscopic quantities extracted from the finite-temperature effective potential: the strength $\alpha$, the inverse duration $\beta/H_*$, the percolation temperature $T_p$, and the bubble wall velocity $v_w$~\cite{Espinosa:2010hh, Grojean:2006bp}. 
The parameter $\alpha$ measures the released vacuum energy relative to the radiation density, while $\beta/H_*$ determines the transition timescale. 
$T_p$ marks the temperature at which bubbles of the true vacuum percolate, and $v_w$ controls how the released energy is distributed. We verified that, in our scenario, $T_p \simeq T_n$, the latter being the nucleation temperature which we use, instead of the former, for the sake of simplicity. 
Together, these parameters dictate the amplitude and shape of the resulting GW spectrum.

\section{Phase Transition Results}

\begin{figure}[h!]
\centering
\includegraphics[scale=0.70]{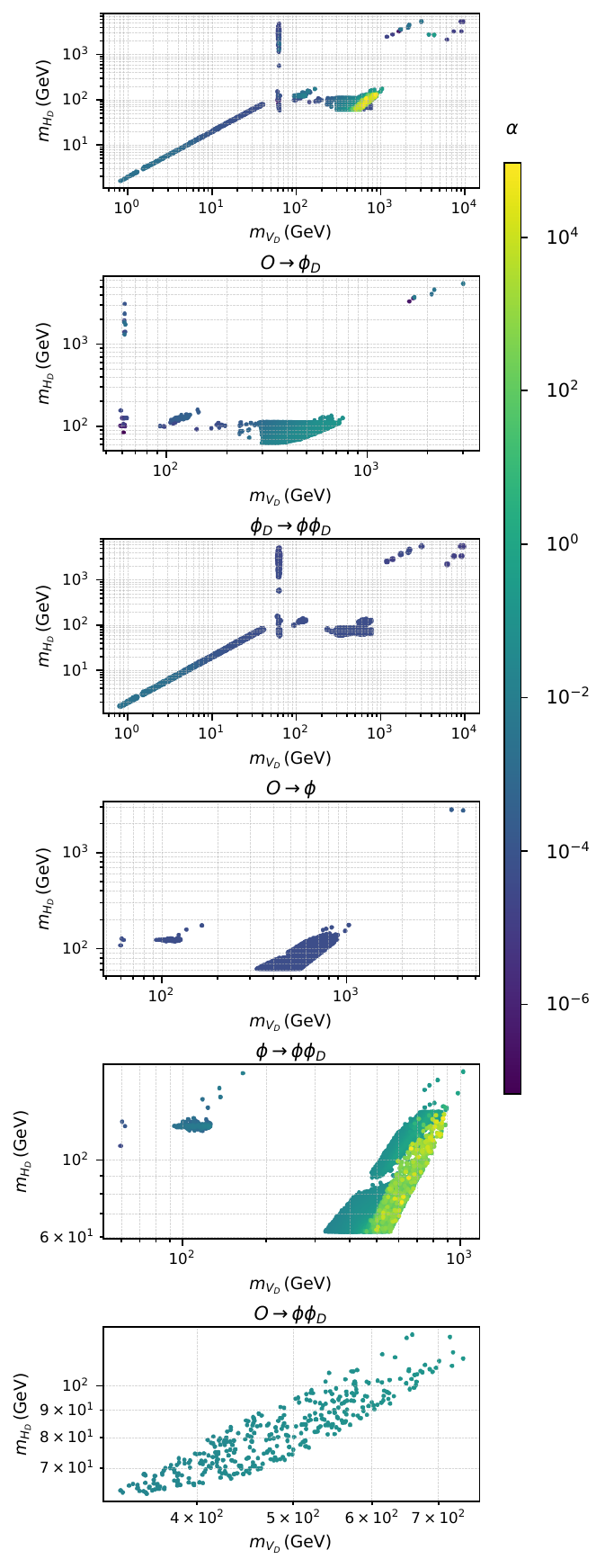}
\caption{Projection of the scanned parameter space onto the $\{m_{V_D}, m_{H_D}\}$ plane. 
The top panel shows the complete set of scan results, while the remaining panels display individual phase transition steps. 
The colour scale in each panel represents the phase transition strength parameter $\alpha$.
}
\label{fig:PhTr}
\end{figure}

We analyse the phase transition dynamics for each viable point in Figure~\ref{fig:summaryscatter} using the \texttt{CosmoTransitions} package~\cite{Wainwright:2011kj}. 
The results are shown in Figure~\ref{fig:PhTr}, which projects the scanned parameter space onto the $\{m_{V_D}, m_{H_D}\}$ plane. 
The colour scale represents the phase transition strength $\alpha$, highlighting the regions that lead to a strong FOPT.

Three characteristic regions appear, consistent with those observed earlier. 
The first corresponds to $m_{V_D} \simeq m_h/2$, the second to $m_{V_D} \simeq m_{H_D}/2$, and the third is a broader region with $m_{V_D} \in [4\times10^2, 10^3]$~GeV and $\vert m_{H_D}-m_h\vert \lesssim m_h/2$. 
Strong FOPTs are mainly found in this last region, where several points also reproduce the observed dark matter relic density. 
A refined scan of this region confirms the presence of configurations with large $\alpha$ values, indicating potentially detectable GW signals.

Among the possible PT patterns, $O \rightarrow \phi_D \rightarrow \phi\phi_D$, $O \rightarrow \phi \rightarrow \phi\phi_D$, and $O \rightarrow \phi\phi_D$, only the double-step transition $O \rightarrow \phi \rightarrow \phi\phi_D$ yields strong transitions with $\alpha \gtrsim 1$. In this notation $O$ represents the origin in field space, a configuration realised at sufficiently high temperatures, $\phi$ and $\phi_D$ represent phases in which only the corresponding scalar gets a VEV, and, finally, $\phi\phi_D$ corresponds to a phase with both scalars acquiring VEVs. The latter is reached at low temperatures.
Transitions beginning directly from the origin tend to be weaker, while the intermediate step $\phi \rightarrow \phi\phi_D$ allows for a larger dark vacuum expectation value $v_D(T_n)$, which enhances the PT strength.

\begin{figure}[h!]
\centering
\includegraphics[scale=0.50]{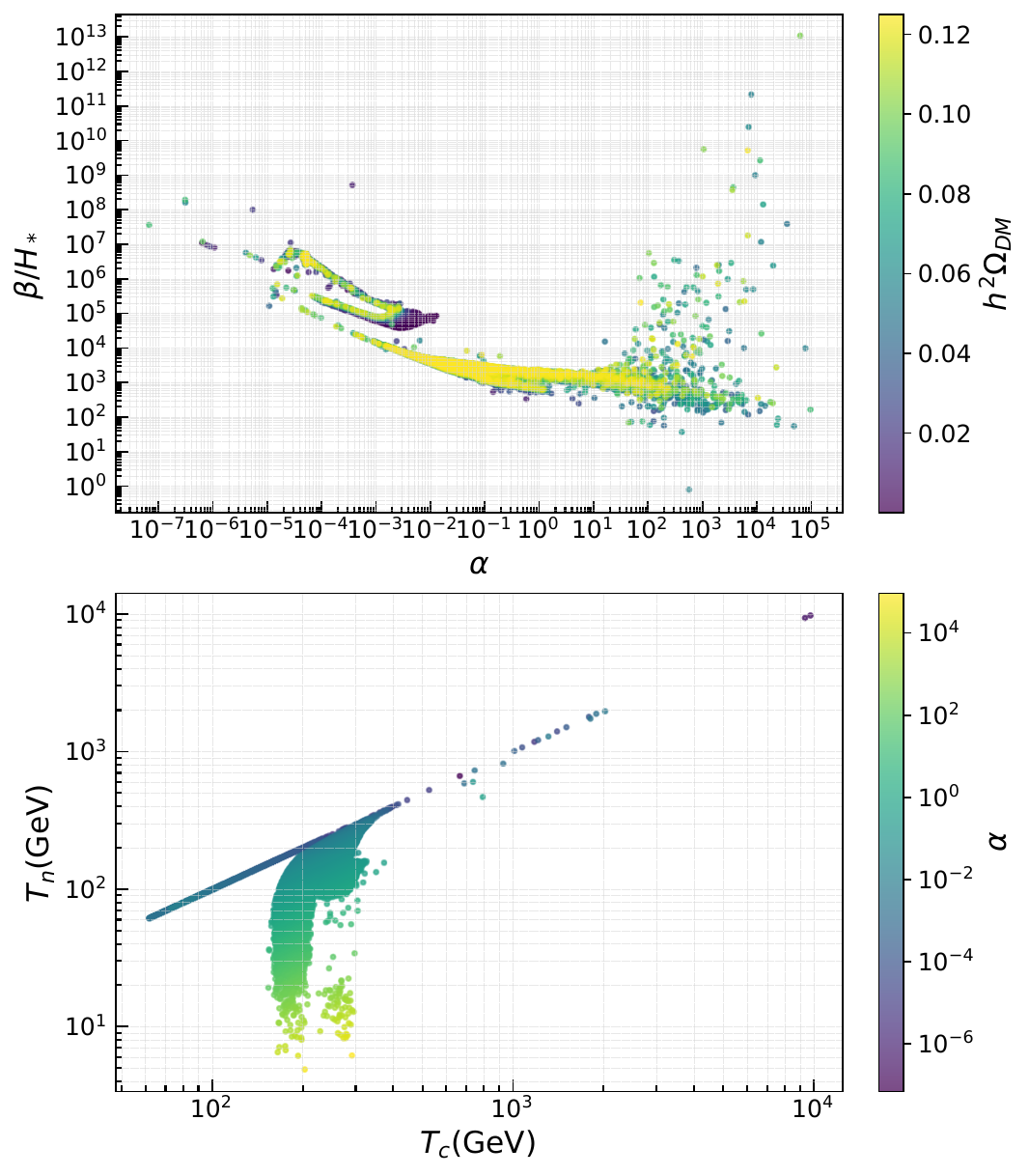}
\caption{The top panel presents how the phase-transition strength $\alpha$ varies with the inverse duration parameter $\beta/H_*,$ where the colour scale reflects the predicted DM abundance. 
In the bottom panel, the connection between the critical temperature $T_c$ and the nucleation temperature $T_n$ is shown, with the colour map representing the corresponding values of $\alpha$ and highlighting how the transition strength evolves with temperature.
}
\label{fig:PhTr2}
\end{figure}

Overall, the strongest FOPTs are associated with dark gauge boson masses in the few-hundred-GeV range, moderate mass splittings between the Higgs and dark Higgs, and small Higgs dark--Higgs mixing angles compatible with collider bounds. 
These regions therefore represent the most promising scenarios for generating GW signals detectable by future observatories.
Figure~\ref{fig:PhTr2} (top panel) illustrates the correlation among the PT parameters $\alpha$ and $\beta/H_*$. 
As expected, we observe that slower transitions, corresponding to smaller $\beta/H_*$ values, are stronger and yield larger $\alpha$. 
In this model, the DM abundance shows no clear correlation with the PT strength, as the observed relic density can be achieved over a broad range of $\alpha$ values. 

The bottom panel of Figure~\ref{fig:PhTr2} displays the critical temperature $T_c$ and the nucleation temperature $T_n$, with the colour scale indicating the PT strength $\alpha$. 
Points lying close to the diagonal line $T_n \simeq T_c$ correspond to transitions that occur almost immediately after the new minimum becomes energetically favoured, resulting in smaller $\alpha$ values. 
Conversely, larger deviations between $T_n$ and $T_c$ indicate slower transitions with greater supercooling, which lead to stronger PTs and hence larger $\alpha$. 
This confirms the negative correlation between $\alpha$ and the temperature ratio $T_n/T_c$.

\section{Gravitational Waves}

An isolated expanding bubble cannot generate GWs, as its spherical symmetry implies a vanishing quadrupole moment. 
A stochastic GW background arises only when multiple bubbles of the true vacuum collide, breaking this symmetry~\cite{Kamionkowski:1993fg}. 
In a thermal plasma, additional GW sources emerge from plasma perturbations: acoustic (sound) waves~\cite{Hogan:1986qda, Hindmarsh:2013xza} and magnetohydrodynamic (MHD) turbulence~\cite{Caprini:2009yp}. 
Subsonic and supersonic bubbles generate compression and rarefaction waves in the plasma, and when these acoustic fronts overlap, the resulting shear stresses in the fluid produce gravitational radiation. 
Moreover, bubble collisions can induce turbulent motion in the plasma, giving rise to a further stochastic GW component.

The total GW power spectrum produced by a FOPT can be expressed as the sum of three main contributions~\cite{Caprini:2015zlo}:
\begin{equation}
    h^2\Omega_{\text{GW}} \simeq 
    h^2\Omega_{\text{col}} + 
    h^2\Omega_{\text{sw}} + 
    h^2\Omega_{\text{turb}}.
\end{equation}
In our model, the dominant contribution originates from sound waves in the plasma~\cite{Caprini:2019egz,Hindmarsh:2017gnf, Schmitz:2020rag}, leading to
\begin{equation}
    h^2\Omega_{\text{GW}}(f) \simeq 
    h^2\Omega_{\text{sw}}(f) = 
    h^2\Omega_{\text{sw}}^{\text{peak}} S_{\text{sw}}(f),
\end{equation}
where
\begin{equation}
\begin{split}
h^2\Omega_{\text{sw}}^{\text{peak}} &= 
1.23\times 10^{-6}
\left(\frac{H_*}{\beta}\right)
\left(\frac{\kappa_{\text{sw}}\alpha}{1+\alpha}\right)^2
\left(\frac{100}{g_*}\right)^{1/3}
v_w \Upsilon, \\[2mm]
S_{\text{sw}}(f) &= 
\left(\frac{f}{f_{\text{sw}}}\right)^3
\left(\frac{7}{4 + 3(f/f_{\text{sw}})^2}\right)^{7/2}.
\end{split}
\end{equation}

Here, $\kappa_{\text{sw}}$ denotes the efficiency factor describing how much vacuum energy is converted into bulk fluid motion, and $f_{\text{sw}}$ is the frequency corresponding to the peak of the sound-wave contribution. 
The suppression factor $\Upsilon$ accounts for the finite duration of the sound-wave period in a radiation-dominated Universe~\cite{Guo:2020grp}:
\begin{equation}
\Upsilon = 1 - \frac{1}{\sqrt{2\tau_{\text{sh}}H + 1}},
\end{equation}
where $\tau_{\text{sh}}$ is the typical lifetime of the sound-wave source. 
In the limit $\tau_{\text{sh}}H \ll 1$, this factor reduces to $\Upsilon \simeq \tau_{\text{sh}}H$.

\begin{figure}[h!]
\centering
\includegraphics[scale=0.34]{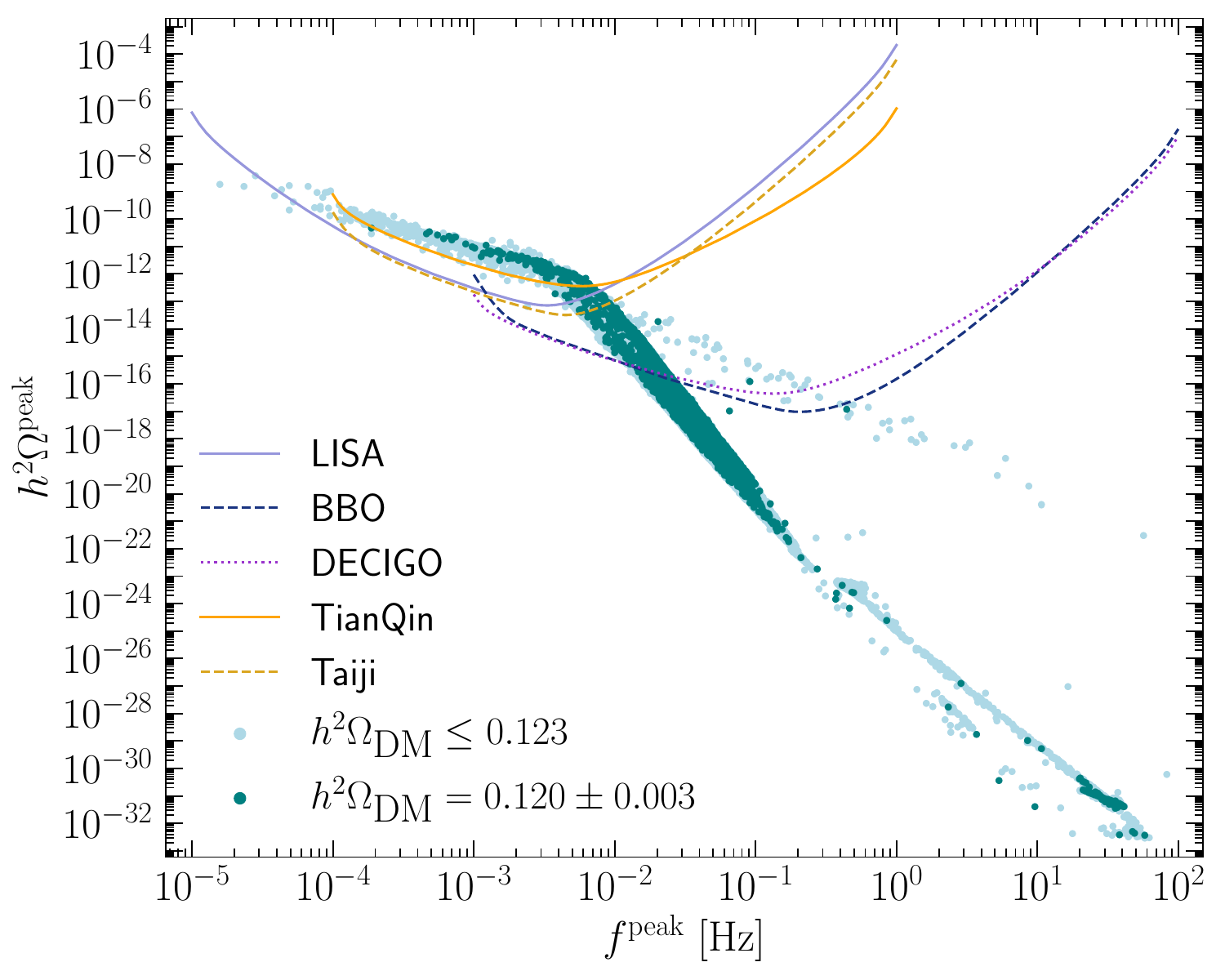}
\caption{Peak GW energy density $h^2\Omega^{\text{peak}}_{\text{GW}}$ versus peak frequency $f^{\text{peak}}$. Darker points indicate parameter sets consistent with the observed dark matter relic density. Power-law integrated sensitivity curves for LISA, DECIGO, BBO, TianQin, and Taiji are shown.}
\label{fig:GWs}
\end{figure}

Figure~\ref{fig:GWs} presents the peak amplitude of the GW power spectrum, $h^2\Omega^{\text{peak}}_\text{GW}$, generated by FOPTs, along with the corresponding peak frequency $f^{\text{peak}}$. 
Darker points indicate parameter configurations that yield the correct DM relic abundance. 
The power-law integrated sensitivity curves~\cite{Thrane:2013oya} for LISA, DECIGO, BBO, TianQin, and Taiji are shown, assuming an observation time of four years and a signal-to-noise ratio of 10. 
A significant fraction of points lie within the reach of these future GW observatories, many of which also reproduce the observed DM abundance. 
The dark points primarily originate from the refined scan focused on strong FOPTs, explaining why configurations with weaker GW signals ($h^2\Omega^{\text{peak}}_\text{GW} \lesssim 10^{-22}$), obtained from the broader scan, are less populated, whereas the refined scan predominantly yields points with $h^2\Omega^{\text{peak}}_\text{GW} \gtrsim 10^{-22}$.

\section{Conclusions}

We have investigated a minimal dark sector extension of the SM that introduces an additional $SU(2)_D$ gauge symmetry. 
In this setup, the SM fields remain neutral under the new gauge group, while the dark scalar doublet interacts with the SM Higgs doublet through a portal term in the scalar potential. 
This interaction leads to mixing between the Higgs and the dark scalar once both fields acquire vacuum expectation values. 
A custodial symmetry in the dark scalar sector naturally stabilises the resulting dark gauge bosons, rendering them suitable dark matter candidates. 

By confronting the model with up-to-date collider, cosmological, and astrophysical data, we identified the parameter regions consistent with all existing bounds. 
Within these regions, the model can account for the observed dark matter relic abundance and simultaneously predict a strong first-order phase transition in the early Universe. 
The corresponding stochastic GW signal is found to be within the sensitivity reach of forthcoming space-based interferometers such as LISA, DECIGO, BBO, TianQin, and Taiji. 

These findings illustrate the complementarity between GW observations and conventional probes of new physics, showing that GW experiments can play a crucial role in testing and constraining dark-sector extensions of the SM.
 
\section*{Acknowledgments}
NB is supported by the National Natural Science Foundation of China (Grant No.~12475105). 
LP’s work is supported by ICSC – Centro Nazionale di Ricerca in High Performance Computing, Big Data and Quantum Computing, funded by the European Union – NextGenerationEU. 
LDR’s work has been funded by the European Union – Next Generation EU through the research grant number P2022Z4P4B \emph{``SOPHYA - Sustainable Optimised PHYsics Algorithms: fundamental physics to build an advanced society''} under the program PRIN 2022 PNRR of the Italian \emph{Ministero dell’Università e Ricerca} (MUR) and by the research grant number 20227S3M3B \emph{``Bubble Dynamics in Cosmological Phase Transitions''} under the program PRIN 2022 of the Italian \emph{Ministero dell’Università e Ricerca} (MUR).

\bibliographystyle{unsrt}
\bibliography{biblio}

\end{document}